\documentclass[twocolumn,trackchanges]{aastex631}

\def\cwt#1 {{\textcolor{black}{#1}}\ }

\begin{document}

\title{Investigating Little Red Dots with UV Excess: Are They the High-Redshift Siblings of Blue Hot DOGs?}

\author[0009-0009-6319-0888]{Lulu Bao}
\affiliation{University of Chinese Academy of Sciences, Beijing 100049, China}

\author[0000-0002-9390-9672]{Chao-Wei Tsai}
\affiliation{National Astronomical Observatories, Chinese Academy of Sciences, 20A Datun Road, Beijing 100101, China}
\affiliation{Institute for Frontiers in Astronomy and Astrophysics, Beijing Normal University, Beijing 102206, China} \affiliation{University of Chinese Academy of Sciences, Beijing 100049, China}
\email{cwtsai@nao.cas.cn}

\author[0000-0001-7808-3756]{Jingwen Wu}
\affiliation{University of Chinese Academy of Sciences, Beijing 100049, China}
\affiliation{National Astronomical Observatories, Chinese Academy of Sciences, 20A Datun Road, Beijing 100101, China}
\email{jingwen@ucas.ac.cn}

\author[0000-0002-2504-2421]{Tao Wang}
\affiliation{School of Astronomy and Space Science, Nanjing University, Nanjing, China}
\affiliation{Key Laboratory of Modern Astronomy and Astrophysics, Nanjing University, Ministry of Education, Nanjing, China}

\author[0000-0003-4007-5771]{Guodong Li}
\affiliation{Kavli Institute for Astronomy and Astrophysics, Peking University, Beijing 100871, People's Republic of China}
\affiliation{National Astronomical Observatories, Chinese Academy of Sciences, 20A Datun Road, Beijing 100101, China}

\author[0000-0002-9508-3667]{Roberto J. Assef}
\affiliation{Instituto de Estudios Astrof\'isicos, Facultad de Ingenier\'ia y Ciencias, Universidad Diego Portales, Av. Ej\'ercito Libertador 441, Santiago, Chile}

\author[0000-0003-0699-6083]{Tanio Diaz-Santos}
\affiliation{Institute of Astrophysics, Foundation for Research and Technology - Hellas (FORTH), Heraklion 70013, Greece}
\affiliation{School of Sciences, European University Cyprus, Diogenes Street, Engomi 1516, Nicosia, Cyprus}

\author{Peter R. M. Eisenhardt}
\affiliation{Jet Propulsion Laboratory, California Institute of Technology, 4800 Oak Grove Drive, Pasadena, CA 91109, USA}

\author[0000-0003-2686-9241]{Daniel Stern}
\affiliation{Jet Propulsion Laboratory, California Institute of Technology, 4800 Oak Grove Drive, Pasadena, CA 91109, USA}

\author[0000-0001-7489-5167]{Andrew W. Blain}
\affiliation{School of Physics and Astronomy, University of Leicester, University Road, Leicester, LE1 7RH, UK}



\begin{abstract}

Little Red Dots (LRDs), newly identified compact and dusty galaxies with an unexpectedly high number density observed by \emph{JWST}, have an unusual “V-shaped” rest-frame UV to near-infrared spectral energy distribution (SED). A group of hyper-luminous, obscured quasars with excess blue emission, called Blue-excess Hot Dust-Obscured Galaxies (BHDs), also exhibit qualitatively similar SEDs to those of LRDs. They represent a rare population of galaxies hosting supermassive black holes (SMBHs) accreting near the Eddington limit at redshifts $z \sim 1$–4. In this study, we compare their multi-wavelength SEDs to investigate whether LRDs, or a subset of them, could be high-redshift analogs of BHDs. 
Our analysis reveals that despite their similar ``V-shape'' SEDs, LRDs appear to be a different population than BHDs. The ``V-shape'' of BHDs appear at longer wavelengths compared to LRDs due to different selection strategies, suggesting LRDs have much less dust attenuation than typical BHDs. The bluer colors in the rest-frame infrared (continuum) emission of LRDs suggest the absence of hot dust heated by AGN accretion activities. We also argue that the blue excess in LRDs is unlikely from AGN scattered light. \textcolor{black}{The compact morphologies and lower X-ray detection frequencies of LRDs suggest a distinct formation pathway from BHDs – which are thought to be powered by super-Eddington accretion onto central SMBHs following major galaxy mergers.}
\end{abstract}



\section{Introduction} \label{sec:intro}

Surveys by the James Webb Space Telescope (\emph{JWST}) have revolutionized our understanding of galaxy formation in the early universe, including the discovery of over-massive black holes at $z > 4$ (\citealt{matthee2024little}, \citealt{kocevski2024rise}). 
Among these findings, a population of red and compact galaxies referred to as ``little red dots'' (LRDs; \citealt{matthee2024little}, \citealt{greene2024uncover}) are distinguished \textcolor{black}{by their distinct near-infrared (NIR) colors, characterized by red rest-frame optical continua and a blue UV excess—commonly described as ``V-shaped'' SEDs (e.g., \citealt{Labbe2023uncover}; \citealt{kocevski2024rise}; \citealt{williams2024galaxies}).} AGN activity in LRDs is suggested by their red optical colors and broad H$\alpha$ emission lines (\citealt{williams2024galaxies}; \citealt{greene2024uncover}). The presence of such abundant SMBHs, with number densities 1–2 dex higher than pre-\emph{JWST} selected quasars in the first Gyr after the Big Bang \citep{kokorev2024census}, may indicate early anomalous growth mechanisms such as super-Eddington accretion, merger events, or massive black hole seeds \citep{greene2024uncover}.
\textcolor{black}{However, LRDs show weak X-ray (\citealt{yue2024}; \citealt{ananna2024ApJ}; \citealt{Andrea2025}; \citealt{maiolino2405jwst}) and millimeter-band emission (\citealt{akins2024cosmos}; \citealt{Xiao2025}; \citealt{Casey2025}), placing constraints on both the AGN emission mechanisms and the dust and gas content of their host galaxies.}

Hot Dust-Obscured Galaxies (Hot DOGs) are a rare population of galaxies discovered by the Wide-field Infrared Survey Explorer (\emph{WISE}; \citealt{Wright_2010}), predominantly at redshifts $z \sim$ 1--4. These galaxies typically exhibit bolometric luminosities $L_{bol} > 10^{13} L_{\odot}$, with some exceeding $L_{bol} > 10^{14} L_{\odot}$ (\citealt{eisenhardt2012first}; \citealt{wu2012submillimeter}; \citealt{Tsai_2015}; \citealt{assef2015half}). Hot DOGs are characterized by very red mid-infrared colors, with SEDs exhibiting a steep rise from rest-frame 3-4 $\mu$m, \textcolor{black}{peaking around 3–100 $\mu$m, and declining at longer wavelengths \citep{wu2012submillimeter,Tsai_2015}}. SED modeling reveals that this strong infrared emission originates from abundant hot dust surrounding an actively accreting central super massive black hole \citep[SMBH;][]{Tsai_2015}. X-ray observations suggest heavy obscuration towards the Active Galactic Nuclei (AGNs), with column densities reaching the Compton-thick limit (\citealt{stern2014nustar}; \citealt{Assef_2016}; \citealt{vito2018heavy}). Hot DOGs at $z\sim1-4$ are massive galaxies with black hole masses ranging from $10^{8} M_{\odot}$ to $10^{10} M_{\odot}$ (\citealt{wu2018eddington}; \citealt{Li_2024}), likely to be in the “blowout” phase after the major merger of two massive galaxies \citep{sanders1988, hopkins2008cosmological}. The discovery of low-redshift Hot DOGs at  $z < 0.5$ by Li et al. (\citeyear{Li_2023}; \citeyear{li2025lowz}) support the interpretation that Hot DOGs represent a special transition phase of AGN accretion histories.

While Hot DOGs consistently show strong hot dust emission in the rest-frame mid-to-far infrared with significant UV-NIR dispersion \citep{Tsai_2015}, a subset exhibit a rest-frame UV excess and are classified as Blue-excess Hot DOGs (BHDs; \citealt{Assef_2016}). In addition to the SED component of a fully obscured AGN, the blue excess of BHDs is best fitted with a second, unobscured AGN component that contributes only $\sim1\%$ of the bolometric luminosity of the obscured component \citep{Assef_2016}.
\textcolor{black}{Subsequent UV polarization observations of selected BHDs confirmed their UV excess originates from scattered AGN light (\citealt{Assef_2022}; \citealt{assef2025}). BHDs may represent Hot DOGs transitioning from a fully obscured AGN phase to a UV-excess phase where the central AGN leaks radiation \citep{Assef_2020,Assef_2022}. Updated statistics indicate up to 27\% of spectroscopically confirmed Hot DOGs can be classified as BHDs \citep{Li_2024}.}

\textcolor{black}{Both BHDs and LRDs exhibit ``V-shaped'' SEDs. For LRDs, the red emission is modeled with a reddened AGN, while the blue rest-frame UV continuum originates from either star formation or scattered light from the obscured AGN (\citealt{Labbe2023uncover}; \citealt{greene2024uncover}; \citealt{leung2024exploring}; \citealt{ma2025}). Notably, the scattered-light scenario shows direct parallels to the physical modeling of BHDs.} 
\textcolor{black}{\citet{Noboriguchi2023} propose an evolutionary sequence connecting LRDs, blue-excess dust-obscured galaxies (BluDOGs; \citealt{Noboriguchi2019}), and BHDs. These systems may represent post-merger galaxies undergoing dusty outflows at different cosmic epochs.}
\textcolor{black}{Alternative models including a single reddened AGN with its extinction curve or a massive, optically thick envelope surrounding the central AGN can also reproduce the ``V-shaped'' feature in LRDs \citep{lizhengrong2025,kido2025}.} 
\textcolor{black}{Beyond their SED characteristics, broad-line LRDs have estimated SMBH masses ranging from $10^{6}-10^{9}\,M_{\odot}$ \citep{greene2024uncover, matthee2024little}, assuming the broad emission lines originate from the broad-line region of a putative AGN. These masses imply that the SMBHs are overmassive relative to their host galaxies \citep{greene2024uncover}. This characteristic is shared with extremely red quasars and Hot DOGs \citep{assef2015half,chen-yu-ching2025}.}
Estimates from broad emission lines would indicate SMBHs in LRDs accrete at sub-Eddington levels, with Eddington ratios (typically $<$0.4) lower than those in Hot DOGs (\citealt{greene2024uncover}; \citealt{matthee2024little}). However, blue-shifted Balmer absorption lines in $\sim$20\% of LRDs reveal gas-rich environments \citep{lin2024spectroscopic}, suggesting possible super-Eddington accretion \citep{inayoshi2024extremely}. 

Given the similar ``V-shaped'' SEDs, possible shared physical scenarios, and a potential evolutionary relation between LRDs and BHDs, we evaluate whether BHDs constitute lower-redshift analogs of LRDs.
This paper is organized as follows: Section 2 introduces the samples of LRDs and BHDs in this work. In Section 3, we examine the differences of UV to NIR SEDs between these two populations. Section 4 discusses selection effects, the properties of SMBHs, and compare the number densities of LRDs and BHDs. We present our conclusions in Section 5. In this paper, we use the AB magnitude system and adopt a flat $\Lambda$CDM cosmology with $H_0$ = 70 km s$^{–1}$ Mpc$^{–1}$, and $\Omega_{M} = 0.3$ \citep{Hinshaw2013}.

\section{Sample selection and Analysis}

\subsection{LRD sample selection and SED collection}
We utilized the photometric data and stacked SED of LRDs from Akins et al. (\citeyear{akins2024cosmos}; hereafter Akins24). \textcolor{black}{This study selected 434 sources from the COSMOS-Web program \citep{Casey2023} and the Public Release IMaging for Extragalactic Research (PRIMER) survey \citep{dunlop2021primer}.} The selection criteria for LRDs in this paper include a compactness metric in F444W, $0.5 < {f_{\rm F444W}(d=0''.2)}/{f_{\rm F444W}(d=0''.5)} < 0.7$, and a red F277W–F444W color, $m_{\rm F277W} - m_{\rm F444W} > 1.5$. This single-color selection criterion for LRDs differs from the other LRD studies using red rest optical $+$ blue UV colors (\citealt{Labbe2023uncover}; \citealt{kocevski2024rise}), and is more similar to the selection of extremely red objects (EROs) \citep{akins2024cosmos}. \textcolor{black}{Notably, the MIRI F1800W imaging of the LRD samples in the PRIMER survey yields marginal or non-detections.}


To evaluate the difference in SEDs between sources selected by Akins24 and those selected by double-color criteria, we picked 76 ``V-shape'' Akins24 sources which satisfy the ``\emph{red2}'' criteria proposed by \citet{Labbe2023uncover}, namely $m_{\rm F150W} - m_{\rm F200W} < 0.8$, $m_{\rm F277W} - m_{\rm F356W} > 0.7$, and $m_{\rm F277W} - m_{\rm F444W} > 1.0$. We also collected the photometry of 260 LRDs from \citet{kokorev2024census}, selected using similar ``V-shape'' selection criteria as \citet{Labbe2023uncover} from the Cosmic Evolution Early Release Science Survey (CEERS; \citealt{Finkelstein2022ceers}), PRIMER, and other programs covering GOODS-S. 
The Akins ``V-shape'' sub-sample and the catalog in \citet{kokorev2024census} have 11 overlapping sources in the COSMOS field. Therefore, in this paper, we focus on these 326 LRDs for SED analysis and comparison, all of which exhibit a ``V-shape'' feature from rest-frame UV to optical.
Additionally, we included six sources detected in both the MIRI/F770W and F1800W bands from \citet{leung2024exploring}, which provides complete photometric coverage from 1–18 $\mu$m using NIRCam and MIRI imaging of 95 LRDs in the UDS and COSMOS fields.

The stacked SED of all LRDs from Akins24 is represented in the top panel of Fig.~\ref{1}. The rest-frame luminosities of the stacked photometry for the “V-shape” sample, \textcolor{black}{derived using photometric redshifts from \citet{kokorev2024census} and Akins24}, along with the 1$\sigma$ scatter, are shown as dark blue dots in \textcolor{black}{eight NIRCam bands (F090W, F115W, F150W, F200W, F277W, F356W, F410W, and F444W) and two MIRI bands (F770W and F1800W)}. The luminosities of stacked photometry of these sub-samples are consistent (within 1 $\sigma$) with the SED of the Akins24 full sample, showing a drop around 0.3 $\mu$m as a result of the ``V-shape'' selection criterion. Notably, the stacked SED of LRDs beyond the observed 18$\,\mu$m in Fig.~\ref{1} is based on the upper limits from non-detections in \emph{Spitzer}/MIPS 24$\,\mu$m, JCMT/SCUBA-2 850$\,\mu$m, ALMA 1.2 and 2.0 mm imaging, and MeerKAT/VLA 1.3 and 3 GHz imaging, following the assumption of multi-temperature dust emission ranging from 30-1500 K \citep{akins2024cosmos}. 

\subsection{Hot DOG SEDs}
\begin{figure*}[!htbp]
    \centering
    \includegraphics[width=1.0\linewidth]{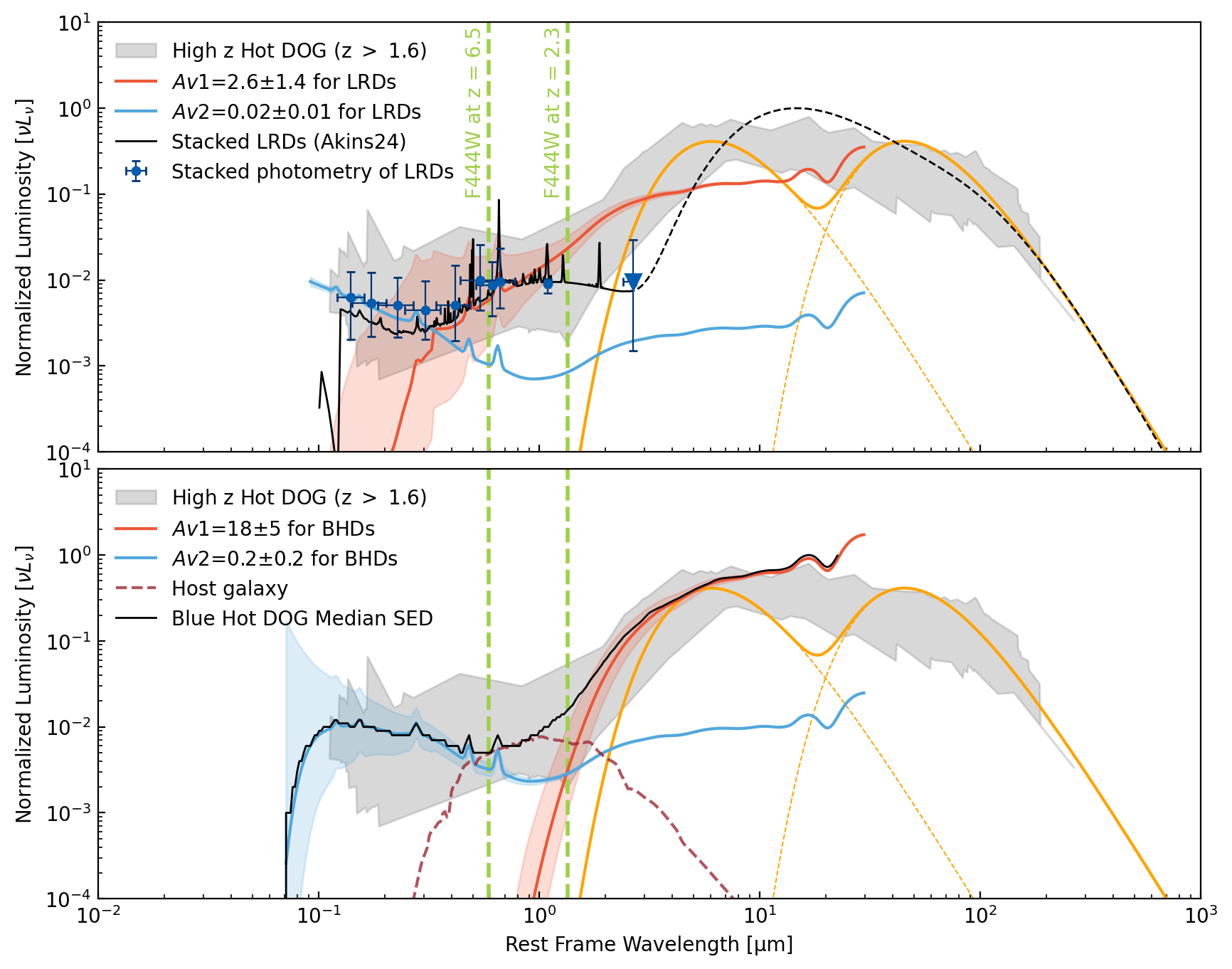}\\ 
    \caption{Comparison between the normalized SEDs of LRDs (upper panel) and BHDs (lower panel). The grey-shaded region represents the distribution of Hot DOGs photometry at $z > 1.6$. \textcolor{black}{The orange lines show dust models with temperatures of 450 K and 60 K and emissivity index $\beta = 1.5$, characterizing the radial slope of the dust density in the single power-law (SPL) model of \citet{Tsai_2015}.} The vertical green dashed lines mark the locations of NIRCam/F444W at the median redshifts of BHDs ($z = 2.3$) and LRDs ($z = 6.5$). 
    \textbf{Top}: 
    The black curve represents the stacked SED of LRDs from Akins24, while the dashed line extending beyond observed 18 $\mu$m shows multi-temperature dust emission assumed by weak far-infrared observations. The blue and red curves with shaded regions correspond to AGN templates with dust attenuation values of A$_{\rm V} \sim$ 0.02 $\pm$ 0.01 mag and 2.5 $\pm$ 1.4 mag, respectively. The dark-blue points indicate the stacked photometry of the compiled 326 “V-shape” LRD sample from Akins24 and \citet{kokorev2024census}. The black-blue triangle represents the stacked MIRI F1800W photometry of 6 MIRI-observed Akins24 sources in the ``V-shape'' sub-sample, shown as upper limits due to non-detection in all ``V-shape'' sub-samples.
    \textbf{Bottom}: 
    The black curve represents the median-stacked SED of BHDs. The blue and red curves, along with their shaded regions, correspond to AGN templates from \citet{Assef_2010}, modeled with dust attenuation values of $A_{\rm V} \sim$ 0.2 $\pm$ 0.2 mag and 18 $\pm$ 5 mag, respectively. The brown dashed line shows the contribution of the host galaxy.}
    \label{1}
\end{figure*}

Hot DOGs are selected by their red mid-infrared colors using the ``W1W2-dropout'' selection criteria \citep{eisenhardt2012first}. The ``core sample'' of Hot DOGs reported in \citet{assef2015half} is defined using slightly more stringent criteria, and includes \textcolor{black}{96 spectroscopically} confirmed W1W2-dropouts at $z > 1$  with W4 $< 7.2$. 
As described earlier, BHDs are Hot DOGs with blue emission excess. Their SEDs are better fit with an extra unobscured AGN component, attributed to the scattering of the blue emission from the obscured AGN \citep{Assef_2016, Assef_2020, Assef_2022}. 
\citet{Li_2024} found that 27\% of the Hot DOGs in the core sample have a mildly reddened AGN component contributing $>$ 50\% of the observed light at rest-frame wavelengths $<$ 1 $\micron$, and hence could be classified as BHDs. 
\textcolor{black}{This fraction is affected by selection bias, as BHDs are easier to obtain secure redshifts among Hot DOGs due to their brighter rest-frame optical emission.}
We note that in terms of bolometric luminosity, the scattered light contributes less than 1\% \citep{Assef_2016, Assef_2020, Assef_2022}.  
In Fig.~\ref{1}, the solid black line in the bottom panel shows the \textcolor{black}{stacked rest-frame SED of BHDs after normalization with respect to their bolometric luminosities, derived from the 26 BHDs in the core sample presented in \citet{Li_2024}}. The gray shaded area represents the normalized SED range for Hot DOG samples at $z > 1.6$ (Tsai et al. in prep).

\section{SED Analysis}

The lower redshifts of the spectroscopically confirmed LRDs from \citet{greene2024uncover} overlaps with the redshift of the most distant Hot DOGs discovered so far (\citealt{Tsai_2015}; \citealt{diaz-santos18}). 
\textcolor{black}{The characteristic ``V-shape'' SEDs observed in both LRDs and BHDs motivate an investigation into whether BHD selection criteria could also identify LRDs, and more broadly, whether the two populations share a common physical origin. To this end, we compare their SEDs and other properties.} 

\begin{figure*}[!htbp]
    \centering
    \includegraphics[width=1.0\linewidth]{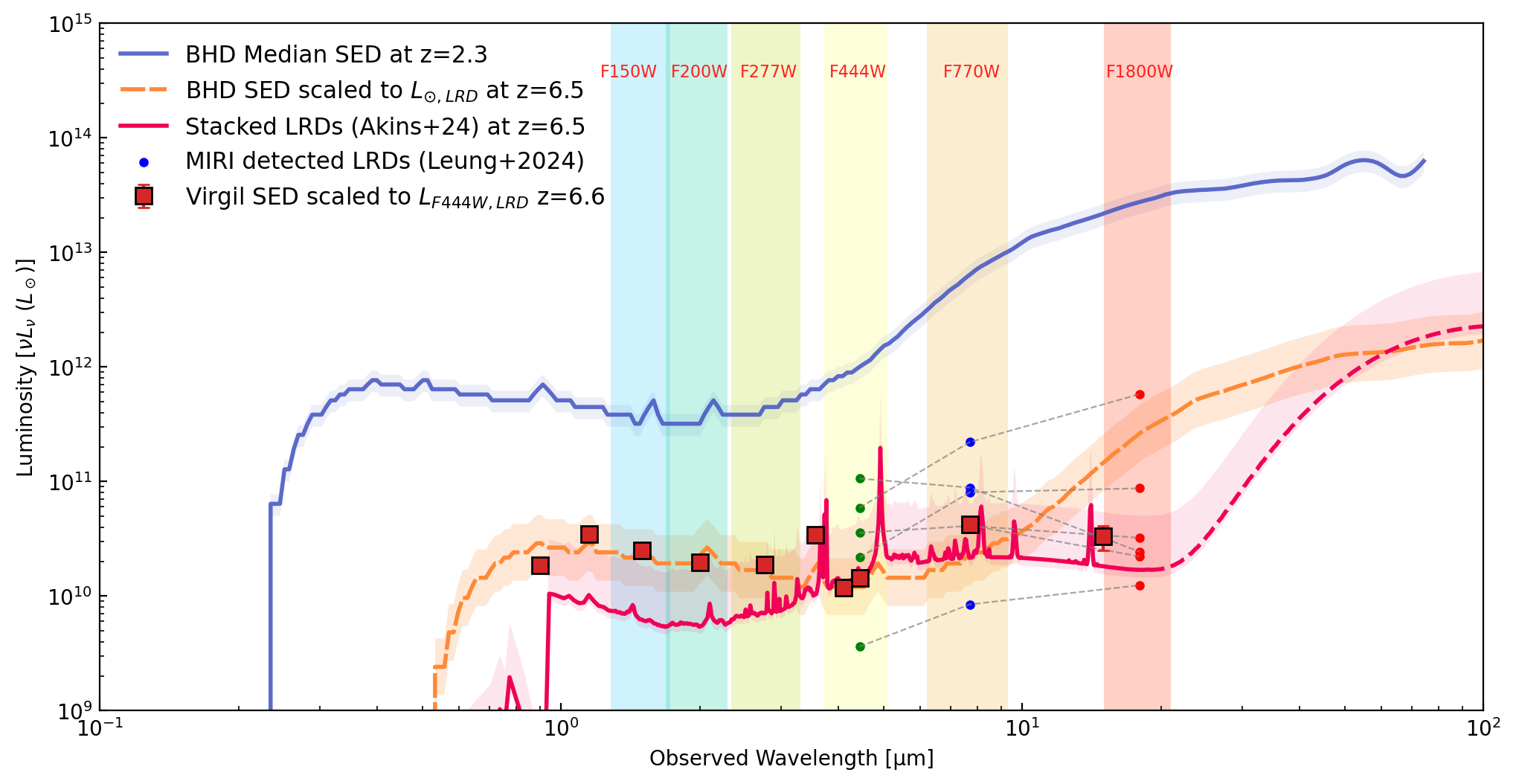}\\
    \caption{The observed-frame SEDs of LRDs and BHDs. The SEDs of BHDs at $z=2.3$ (blue solid curve) are scaled to their median bolometric luminosity $L_{\mathrm{bol,med}} = 6.4\times10^{13} L_{\odot}$. The stacked SEDs of the Akins24 sample at $z=6.5$ and scaled to $L_{\mathrm{bol,med}} = 2.3\times10^{12} L_{\odot}$ from \citet{akins2024cosmos}, are represented by the red curve. The orange curve shows the SED of BHDs at z = 6.5, scaled to the luminosity for the typical SMBH mass of $\sim 10^8 M_{\odot}$ in LRDs \citep{greene2024uncover} and assuming the median accretion rate of BHDs. The small dots in green, blue, and red represent the photometry at F444W, F770W, and F1800W, respectively, for the five LRDs detected in both MIRI bands by \citet{leung2024exploring}. \textcolor{black}{The red squares mark the photometry of \emph{Virgil} \citep{Rinaldi2025}, a compact and MIRI-red AGN at $z=6.6$, with photometry scaled to match the NIRCam/F444W luminosity of the stacked Akins24 LRDs.}}
    \label{2}
\end{figure*}

Fig.~\ref{1} shows normalized SEDs of LRDs and BHDs, modeled using linear combinations of AGN and host galaxy SED templates from \citet{Assef_2010}. In this model, the optical-to-NIR emission is attributed to a reddened AGN, while the UV excess arises from a lightly obscured or unobscured AGN component\textcolor{black}{, with additional contributions from the host galaxy in some cases (shown by the brown dashed line in lower panel of Fig.~\ref{1})}. The SED of BHDs (bottom panel) is shown with two AGN components: a heavily obscured AGN ($A_{\rm V1,med}=18$; \citealt{assef2015half}) and an unobscured counterpart ($A_{\rm V2,med}=0.2$; \citealt{Li_2024}). \textcolor{black}{This fitting methodology (see Section 3.1 of \citealt{Li_2024} for details), previously applied to BHDs \citep{Assef_2016, Assef_2020, Li_2024},} is also adopted here to model the SEDs of ``V-shape'' LRDs with more than three-band detections (top panel). The best-fit model for LRDs consists of a moderately obscured AGN ($A_{\rm V1} = 2.6 \pm 1.4$) dominating the optical emission and a nearly unobscured AGN component ($A_{\rm V2} = 0.02 \pm 0.01$) responsible for the blue excess. The fitting results of LRDs is consistent with previous SED-based studies of LRDs, which derive $A_{\rm V} \sim 3$ for the reddened AGN component \citep[e.g.,][]{Labbe2023uncover, greene2024uncover, leung2024exploring}.

The vertical green dashed lines shown in Fig.~\ref{1} mark the locations of the NIRCam/F444W filter at redshifts $z = 6.5$ and $z = 2.3$, corresponding to the median redshift of the Akins24 sample and BHDs, respectively. 
At $z = 6.5$, the F444W filter captures the brightest emission of LRDs, while it corresponds to the valley bottom of “V-shape” SED for BHDs. On the contrary, BHDs at $z=6.5$ are expected to be bright in F1800W band due to the dominant dust emission, whereas \textcolor{black}{LRDs are mostly faint in F1800W and remain undetected in the millimeter bands (\citealt{Xiao2025}; \citealt{Casey2025}), suggesting a lack of significant dust emission above the detection threshold.}
The difference in the ``V-shape'' valley points between LRDs and BHDs indicate that the the \emph{red2} selection criteria cannot identify BHD candidates with similar redshifts as LRDs but may preferentially select BHDs at lower redshifts ($z \sim$ 2--3).

Fig.~\ref{2} highlights the significant differences in bolometric luminosity between the observed-frame SEDs of BHDs at $z = 2.3$ and the Akins24 sample at a median redshift of $z_{\rm med} = 6.5$. We also plot the MIRI detections of selected LRDs from \citet{leung2024exploring}. 
The red and blue curves are scaled to the median bolometric luminosities of $L_{\rm bol,med} = 6.4 \times 10^{13} L_{\odot}$ for BHDs and $L_{\rm bol,med} = 2.3 \times 10^{12} L_{\odot}$ for LRDs, respectively. 
The orange curve shows the BHD SED scaled to match the bolometric luminosity expected from an AGN with the typical LRD black hole mass ($\sim 10^8\,M_{\odot}$, estimated from H$\alpha$; \citealt{greene2024uncover, matthee2024little}) accreting at the median BHD Eddington ratio ($\sim 0.7$; \citealt{Li_2024}). The shaded regions represent the uncertainty of each curve, calculated using the Median Absolute Deviation (MAD).
Although the \textcolor{black}{scaled luminosity} is comparable to that of LRDs, the resulting SED shows a continuously declining trend across the \emph{JWST}/NIRCam bands, which contrasts with SEDs typically seen in LRDs. This SED inconsistency with typical LRDs makes BHD classification as an LRD unlikely.

\textcolor{black}{The red squares in Fig.~\ref{2} show the photometry of \emph{Virgil}, a dusty, obscured AGN at $z = 6.6$ \citep{Rinaldi2025}, identified through its red MIRI colors \citep{Iani2024}. While its rest-frame UV-to-optical colors exclude it from the LRD selection criteria, this galaxy shares the compact morphology characteristic of LRDs. According to \citet{Noboriguchi2023}, both LRDs at $z > 5$ and UV-excess dust-obscured galaxies (DOGs) at cosmic noon may represent a post-merger outflow phase, with varying covering factors of dusty material, suggesting a potential evolutionary sequence.
\emph{Virgil} may serve as a bridge between LRDs and DOGs. Its SED resembles that of BHDs blueward of F770W, but diverges more significantly from LRDs. Notably, its F770W\,–\,F1500W color is bluer than that of BHDs, indicating a deficiency of hot dust emission around the central AGN.}

\section{Discussion}

\subsection{Selection-induced differences in BHDs/LRDs}

The rest-frame UV to near-infrared colors of LRDs and BHDs reveal notable differences in the SEDs of these two populations. The ``W1W2-dropout'' criterion efficiently selects galaxies with significant dust emission, which contributes approximately 99\% of $L_{\rm bol}$ starting from $\sim$3--4\,$\mu$m. In contrast to typical AGN torus emission models, the SEDs of LRDs observed with \emph{JWST}/MIRI remain relatively flat in the mid-infrared bands \citep{williams2024galaxies, perez2024nature}.
In the full Akins24 sample, no sources are detected at F1800W. In another study, two LRDs are detected in the F1800W band out of a sample of 31 ($\sim$6.5\%) \citep{perez2024nature}, consistent with the detection rate of LRDs in both the F770W and F1800W filters (6/100, 6\%) from MIRI observations \citep{leung2024exploring}. These results collectively confirm the low F1800W detection rate in LRDs. The bluer F770W\,–\,F1800W color ($< 2$ mag) compared to BHDs ($\sim 5$ mag) reveals a deficiency of hot dust, implying less AGN dust heating in these abundant \emph{JWST}-detected infrared sources.

\textcolor{black}{A key selection criterion for LRDs is their compact morphology, which distinguishes them from populations like Hot DOGs that are selected purely by color. This results in a significant size difference: LRDs have a median effective radius of $R_e \sim 100$ pc \citep{akins2024cosmos, kokorev2024census}, while Hot DOGs are much more extended with $R_e \sim 1$--$4$ kpc \citep{Farrah2017,diaz-santos2021}. The morphology of these objects can provide clues to the host galaxy's contribution to their blue emission. While obscured star formation was considered as a possible way to explain the extended UV emission in BHDs \citep{Assef_2020}, high polarization seen in BHDs confirms that the extended UV emission is instead due to light scattered from an outflow \citep{assef2025}. In contrast, for some LRDs, exhibiting a resolved UV morphology is evidence of a contributing host galaxy \citep{pinaldi2024,wangbingjie2024,Chen2025}.}

\subsection{Comparison of UV origins via SED fitting}
\textcolor{black}{BHDs exhibit significant dust attenuation of $\sim$20–60 mag in the rest-frame optical to NIR continuum \citep{assef2015half}, with strong evidence that the observed UV excess originates from scattered light escaping through lightly obscured or unobscured lines of sight \citep{Assef_2022}.}
In contrast, the valley points of the ``V-shape'' feature in LRDs are generally bluer than those of BHDs, as shown in Fig.~\ref{1}.
\textcolor{black}{Adopting the same SED fitting methods as BHDs, our analysis indicates significantly lower attenuation ($A_{\rm V} \, \sim \, 2.6$) toward the reddened AGN component. The luminosity ratio between blue and reddened components indicates scattered light contributes $\lesssim 1\%$ of the total emission, consistent with other estimations from LRD samples (\citealt{labbe2024b}; \citealt{kocevski2024rise}; \citealt{leung2024exploring}). This fraction is comparable to that inferred for BHDs \citep{Assef_2016}.}
\textcolor{black}{Within the framework of the BHD scattering scenario, it is therefore difficult to reconcile how LRDs could yield a comparable scattered-light fraction under such reduced obscuration.}
More likely, the observed blue light in LRDs is not significantly contributed by scattered UV photons from the AGN leaking through a narrow, low-obscuration opening angle. \textcolor{black}{Therefore, we propose that the blue excess contributing to the ``V-shape'' feature in LRDs may not originate primarily from scattered UV photons from a central AGN via a mechanism like that in BHDs.}



A similar conclusion is reached in some other works. For example, the high similarity in the elevated red side of ``V-shape'' SEDs commonly seen in LRDs suggests that the model of a reddened AGN with partial scattered emission cannot explain the observed SEDs of LRDs well (\citealt{setton2024}; \citealt{labbe2024b}).
In addition, the lack of significant photometric variability in the 21 “V-shape” LRDs from \citet{WeiLeong2024} implies an upper limit of $\sim$ 30\% for the total observed UV light attributed to the AGN contribution. At rest-frame optical wavelengths, the unresolved morphology suggests an AGN's dominance in LRDs. However, at shorter wavelengths, approximately 30\% of LRDs exhibit resolved UV morphologies \citep{wangbingjie2024,pinaldi2024,Chen2025} suggesting that the rest-frame UV continuum could be primarily contributed by host galaxies in LRDs. 

\textcolor{black}{Recent observations of a luminous quasar at $z=2.566$, considered a “Big Red Dot” analog of LRDs \citep{Stepney2024}, offer further insight into the scattering scenario. The source lacks the hot dust typically associated with a central torus, suggesting that both obscuration and scattering occur on galaxy scales rather than within a compact nuclear structure. Notably, its UV excess can be explained by only $\sim$0.05\% of the intrinsic AGN light and may also include contributions from star formation in the host galaxy. These results support the interpretation that the blue excess observed in LRDs is unlikely to originate from classical torus-scale scattering mechanisms.}

\begin{figure}[!htbp]
    \centering
    \includegraphics[width=1.0\linewidth]{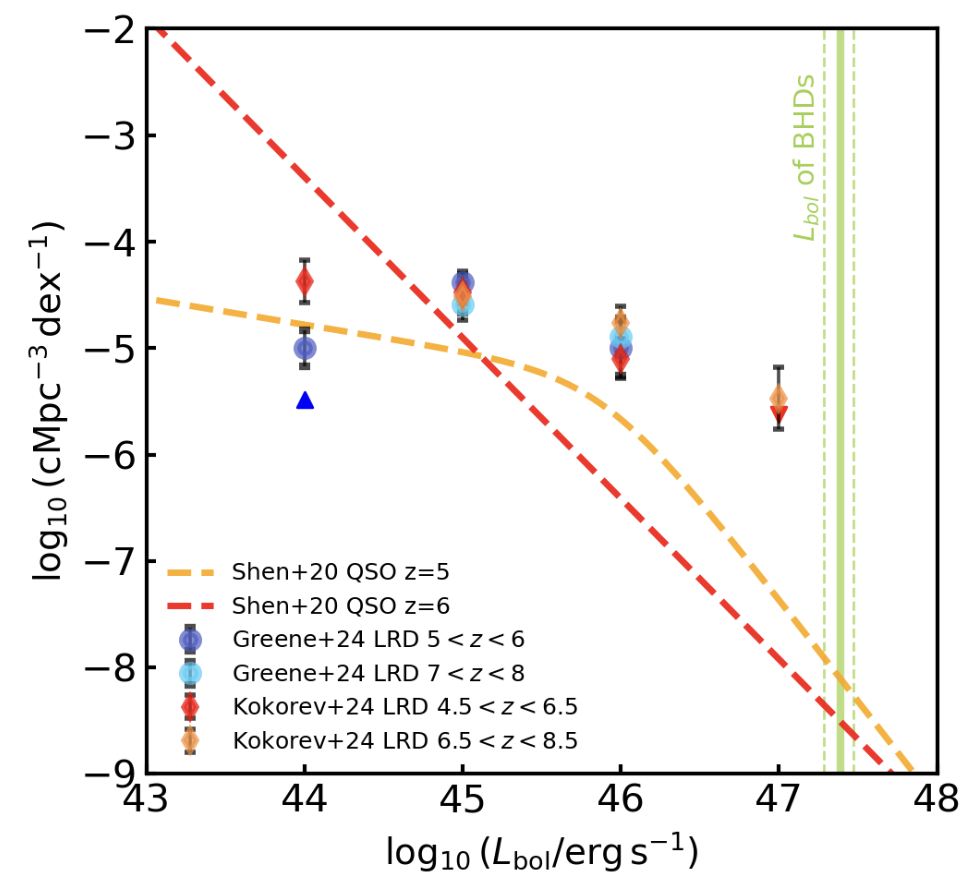}\\ 
    \caption{The bolometric LFs of LRDs in higher and lower redshift bins reported by \citet{kokorev2024census} and \citet{greene2024uncover} are presented in diamonds and points, respectively. The pre-\emph{JWST} bolometric LFs at $z=6$ and $z=5$, derived by \citet{shen2020}, are shown as dashed red and orange lines, respectively. The vertical green solid and dashed curves indicate the median bolometric luminosity of BHDs, with its 1$\sigma$ scatter.
    }
    \label{3}
\end{figure}

\subsection{Number density comparisons}
At $z\sim 2$–4, Hot DOGs exhibit number densities comparable to equally luminous unobscured quasars \citep{assef2015half}, indicating significant obscuration among the most luminous AGN. Assuming similar obscured fractions persist at higher redshifts and consistent BHD/Hot DOG ratios, we estimate the $z\sim 6$–8 BHD number density using bolometric luminosity functions (LFs).
Figure~\ref{3} presents the number density of LRDs from \citet{greene2024uncover} and \citet{kokorev2024census}. For BHDs, the median bolometric luminosity is $L_{bol} \sim 7.8^{+2.8}_{-1.4} \times 10^{13} L_{\odot}$. At $z=6$, BHDs have an estimated number density of $\sim 10^{-8}$--$10^{-9} \, \rm{Mpc}^{-3} \rm{dex}^{-1}$. This low density is primarily attributable to their extreme luminosity. Consequently, BHD-luminosity quasars are highly rare among \emph{JWST}-selected sources.

\textcolor{black}{An evolutionary connection between LRDs and BHDs would predict either the coexistence of BHD-like systems with LRDs at $z > 5$, or LRD analogs appearing at cosmic noon. However, as discussed above, the number density of LRDs steeply declines at $z < 5$, where Hot DOGs begin to emerge. This trend does not rule out the possibility of observational biases.
The currently known MIRI-selected, extremely red AGN \emph{Virgil} is a rare \emph{JWST}-detected source. It exhibits a UV-to-NIR SED similar to that of BHDs, but shows a bluer mid-infrared color. Its rarity suggests that the intense AGN heating mechanism to produce hot dust is rare in the high-redshift Universe. Thus, systematic searches for high-$z$ Hot DOGs are needed in future studies.
Notably, the fraction of “V-shaped” sources among high-$z$ broad-line and compact AGNs ($\sim 2/3$; \citealt{matthee2024little,Hviding2025}) is substantially higher than the fraction of BHDs among Hot DOGs ($\sim$27\%; \citealt{Li_2024}). BHDs are thought to represent a transitional phase in which AGN light begins to leak from heavily obscured cocoon, accompanied by strong UV metal emission lines (e.g., [C\,\textsc{iv}]; \citealt{Assef_2020}). In contrast, the high UV-excess fraction implies weak dust signals and may point to an evolving dust-to-gas ratio over cosmic time.
Searches for LRD analogs in the local Universe by \citet{lin2025local} find lower-limit number densities of $\sim 10^{-10}~{\rm Mpc^{-3}}$, reinforcing the rarity of low-redshift LRDs.
Recent theoretical studies \citep{Inayoshi2025} propose that the first one or two AGN events associated with newly formed seed black holes are observed as LRDs, whose distinctive features fade in subsequent episodes.
While intriguing parallels exist, the evolutionary link between LRDs and BHDs remains speculative and requires further investigation.}

\subsection{SMBHs in LRDs and BHDs: both under Eddington limit accretion?}

\textcolor{black}{The emission lines (e.g., H$\alpha$ and [O~\textsc{iii}]) in Hot DOGs indicate the presence of powerful ionized outflows \citep{wu2018eddington, Jun_2020, Finnerty2020, Vayner2024, Liu2025}.}
Combining the high Eddington ratios and outflow features, these properties indicate that Hot DOGs are undergoing intense accretion and feedback activities, with some potentially surpassing the Eddington accretion limit (\citealt{Tsai_2015}; \citealt{Li_2024}). 
Similar to LRDs, these galaxies are typically faint or undetected in X-ray observations \citep{stern2014nustar}. \textcolor{black}{The general X-ray non-detection of LRDs (\citealt{yue2024}; \citealt{ananna2024ApJ}; \citealt{maiolino2405jwst})} and the $\sim$ 20\% detection rate of absorption features in broad line LRDs are thought to be due to gas-rich environments of AGNs in these galaxies, which may be explained by super-Eddington accretion (\citealt{greene2024uncover}; \citealt{inayoshi2024extremely}). 

We cross-matched the LRD sample with catalogs from \emph{Chandra} deep field surveys (\citealt{nandra2015ApJS}; \citealt{xue2016ApJS}; \citealt{marchesi2016ApJ}; \citealt{luo2017ApJS}; \citealt{kocevski2018ApJS}). Except for two X-ray-detected LRDs reported by \citet{kocevski2024rise}, no other LRDs were identified in these X-ray observations. One of the two X-ray–detected LRDs is at $z=4.66$ and was only marginally detected, with a column density of $N_{\rm H}\gtrsim10^{23}\,\mathrm{cm}^{-2}$, indicating a moderately obscured AGN. Another X-ray LRD is best fitted with column density $N_{\rm H} \sim 10^{22}$ cm$^{-2}$. Both measurements agree with the $E(B-V)/N_{\rm H}$ ratio reported by \citet{maiolino2001dust} \citep{kocevski2024rise}. 

In comparison, 13 out of 44 BHDs have been observed by \emph{Chandra} and \emph{XMM-Newton}, and five are detected with exposure times $>$ 10 ks (\citealt{stern2014nustar}; \citealt{vito2018heavy}; \citealt{Assef_2016}; \citeyear{Assef_2020}). Additionally, one BHD is identified in the eROSITA all-sky survey \citep{erosita} with a detection likelihood $>$ 6. The gas column density of these sources consistently exceeds $10^{23}\, \mathrm{cm^{-2}}$, confirming the presence of heavily obscured AGNs at their centers. 

\textcolor{black}{Compared to BHDs, the non-detection in stacked X-ray images of LRDs provides a stringent constraint (\citealt{yue2024}; \citealt{ananna2024ApJ}), which may reflect a combination of factors, including heavy obscuration, softer X-ray spectra associated with high Eddington ratios, and/or intrinsic X-ray weakness possibly due to a quenched corona \citep{Luo2013}. Theoretical models further support the plausibility of super-Eddington accretion in LRDs \citep{inayoshi2024extremely,Madau2025,Pacucci2024}.} Combined with their near-infrared SEDs, LRDs are unlikely to resemble BHD-like sources, which are speculated to result from major mergers, being nearly completely obscured and undergoing super-Eddington accretion. \textcolor{black}{We note, however, that differences in X-ray exposure depths and the rest-frame energies probed at different redshifts between LRDs and BHDs may affect this comparison.}
\citet{maiolino2405jwst} also mentioned that misclassifications of non-AGN sources, including supernovae or extreme star-forming galaxies, cannot be entirely excluded. 

The unexpectedly high number density of AGN-powered LRDs raises an intriguing puzzle about how did these black holes ($\sim 10^{6-8}$ M$_{\odot}$) form and grow within the first billion years of the universe. To relate the density of LRDs with the observed black hole density at lower redshifts, these abundant SMBHs in LRDs are thought to be rapidly spinning and have radiative efficiencies $>$ 0.1, sustained by prolonged, orderly mass accretion rather than chaotic accretion \citep{inayoshi2024birth}. 
In addition, large black hole seeds, high Eddington accretion rates, and high merger rates may be other explanations. The latter is supported by the large fraction of broad line LRDs showing nearby companions (\citealt{matthee2024little}; \citealt{lin2024spectroscopic}). 
For comparison, Hot DOGs are potentially suffering chaotic accretion after merger events (\citealt{Tsai_2015}; \citealt{Jun_2020}). Hot DOGs at higher redshifts ($z\gtrsim3$) typically host SMBHs with masses $M_{\rm BH} \sim 10^9 M_{\odot}$ (\citealt{Tsai_2015}; \citealt{Li_2024}). These SMBHs require massive black hole seeds, super-Eddington accretion for at least $>$ 25\% of their growth time, and low radiative efficiencies, possibly due to low BH spin \citep{Tsai_2015}. 
The distinct properties between LRDs and BHDs may suggest divergent black hole accretion pathways in the early Universe.


\section{Conclusions}

Motivated by the similarity in their observed properties, particularly the ``V-shape'' feature in SEDs and thus the possible similarity in their galaxy components, we explored the potential connections of the excess blue light between the rare Blue Hot DOGs (BHDs) sample and the newly discovered Little Red Dots (LRDs).

\textcolor{black}{However, LRDs exhibit only faint emission beyond rest-frame NIR, whereas in BHDs the bolometric luminosity is dominated by emission at wavelengths longer than rest-frame 1 $\mu$m. This indicates that their SEDs are fundamentally different.}

BHDs are best fitted with a heavily obscured AGN component ($A_{\rm V} \sim$ 20–60 mag) with slight ($\sim$ 1\%) UV light leakage from the AGN. In contrast, the “V-shape” in LRDs appears at shorter wavelengths due to their lower dust attenuation ($A_{\rm V} \sim 2.6$ mag). Polarimetric observations indicate that UV excess in the BHDs is likely scattered light from an unreddened AGN, which is unlikely to occur in LRDs, given LRDs have a similar fraction of scattered light but much lower dust attenuation compared to BHDs.


BHDs have a higher X-ray detection rate compared to LRDs. As BHDs represent a transitional phase in which light from the accretion disk begins to leak through a previously fully obscuring torus, the rarity of mid-infrared bumps and the low X-ray detection rates in LRDs suggest that they may lack the heating mechanism associated with Eddington-limit accreting BHD-like AGNs in the post-merger phase.

\begin{acknowledgments}
Authors thank Hollis B. Akins and collaborators for kindly providing the photometric data of LRDs from their recent work. This work is supported by NSFC 12588202 and International Partnership Program of the Chinese Academy of Sciences, Grant No. 114A11KYSB20210010.
\end{acknowledgments}

%

\vspace{5mm}





\bibliography{Bao_LRD_and_BHDs}{}
\bibliographystyle{aasjournal}



\end{document}